\documentclass[conference]{IEEEtran}
\IEEEoverridecommandlockouts

\usepackage{cite}
\usepackage{amsmath,amssymb,amsfonts}
\usepackage{algorithmic}
\usepackage{graphicx}
\usepackage{textcomp}
\usepackage[table]{xcolor}
\usepackage{multirow}
\usepackage{pifont}
\usepackage{makecell}
\usepackage{subcaption}
\usepackage{booktabs}
\usepackage{url}
\usepackage{tikz}
\usetikzlibrary{positioning, arrows.meta, shapes.geometric, fit, calc}
\def\BibTeX{{\rm B\kern-.05em{\sc i\kern-.025em b}\kern-.08em
    T\kern-.1667em\lower.7ex\hbox{E}\kern-.125emX}}

\title{LLM can Read Spectrogram: Encoder-Free Speech-Language Modeling}

\author{\IEEEauthorblockN{Ruchao Fan, Yiming Wang$^*$\thanks{* Contributed to the work in 2025 before leaving Microsoft.}, Yuxuan Hu, Bo Ren, Yufei Xia,
Xiaofei Wang, Yao Qian, Shujie Liu, Jinyu Li}
\IEEEauthorblockA{Microsoft, USA}}

\begin{document}
\maketitle
\begin{abstract}
Recent speech-aware large language models (Speech-LLMs) rely on pre-trained speech encoders to convert audio into semantic/acoustic rich representations consumable by LLM.
In this work, instead, we explore: \emph{can an LLM learn to read Mel spectrogram directly without a dedicated speech encoder?}
We propose \textbf{Mel-LLM}, an encoder-free Speech-LLM that feeds lightly pre-processed Mel-spectrogram patches directly into the LLM through a linear projection, allowing the LLM to learn speech-text alignment purely through its own parameters.
We focus on speech understanding tasks, including automatic speech recognition (ASR), spoken QA and audio understanding.
For ASR, we evaluate on the OpenASR Leaderboard public sets and production-level scaling experiments, demonstrating that the encoder-free solution achieves competitive performance with only limited degradation compared to encoder-initialized counterparts. We find that when data is limited, initialization from a multimodal checkpoint (Phi-4-MM) is crucial for maintaining performance. We also present ablation studies suggesting which LLM layers are most involved in speech adaptation. Beyond ASR, we extend Mel-LLM with general speech/audio understanding tasks, revealing an acoustic-semantic trade-off: directly exposing the LLM to Mel-spectrogram input improves paralinguistic and non-ASR acoustic tasks, while knowledge-intensive spoken QA remains more challenging than encoder-anchored systems.
We additionally include a text-to-speech (TTS) proof-of-concept with a next-token VAE decoder, showing that direct Mel generation is possible but still trails stronger latent-diffusion generation.
\end{abstract}

\section{Introduction}
\label{sec:intro}
\vspace{-0.05cm}

The prevailing paradigm for speech large language models (Speech-LLMs)~\cite{FathullahWLJSLG24,ShiJXXZWSZY24,abouelenin2025phi} consists of three components: a pre-trained speech encoder, a modality projector, and a large language model (LLM). The speech encoder, typically a Whisper-style~\cite{radford2023robust} or Conformer-based~\cite{GulatiQCPZYHWZW20} model pre-trained on large-scale ASR data, converts raw audio into high-level speech representations. These are then projected into the LLM's embedding space for downstream tasks such as ASR, translation, instruction following, and spoken QA.

While effective, this reliance on a dedicated speech encoder introduces several limitations. First, the encoder is often large (e.g., 600M+ parameters for Whisper-large) and adds significant computational overhead. Second, the encoder's learned representations may not be optimal for the LLM's internal processing, creating a representational mismatch by late fusion. Third, the encoder becomes a bottleneck for information flow---the LLM can only access speech through the encoder's compressed representations, either semantic or acoustic rich. As a result, recent speech-LLMs adopted dual-encoder design with dedicated training for semantic and acoustic information preservation, such as WavLLM~\cite{hu2024wavllm}, SALMONN~\cite{tang2024salmonn}, and VibeVoice~\cite{peng2025vibevoice}. Such dual-encoder designs complicate the architecture and require strong speech domain knowledge, e.g. from transformer to conformer~\cite{GulatiQCPZYHWZW20} for speech understanding, and VAE choice for speech generation. Recent works in vision-language models have shown that LLMs can directly process raw pixel patches~\cite{fuyu2023} without a vision encoder, suggesting that large language models have sufficient capacity to learn modality-specific processing internally. Inspired by this, we ask: \emph{can an LLM directly read Mel spectrogram?}

In this paper, we propose \textbf{Mel-LLM}, an encoder-free Speech-LLM architecture that removes the pre-trained speech encoder blocks, while optionally retaining lightweight convolution layers for temporal downsampling purpose only. The Mel spectrogram is chunked in time and projected directly into the LLM's embedding space. This design simplifies the Speech-LLM architecture,  and allow the LLM itself learns to interpret these raw spectral features and align them with text. Our experiments suggest that when the LLM is sufficiently large, its lower layers can learn part of the speech-adaptation function normally handled by a dedicated encoder. Removing the main speech encoder blocks enables direct speech-text modeling within the LLM backbone for understanding tasks, and provides a path toward future unified speech generation.

Our contributions are as follows:
\begin{itemize}
    \item We demonstrate that an LLM can learn to perform ASR directly from Mel spectrogram without a dedicated Transformer/Conformer speech encoder, achieving competitive results on the OpenASR leaderboard.
    \item We show that encoder-free models have limited degradation compared to encoder-initialized ones at scale.
    \item We extend the study beyond ASR to speech understanding tasks, showing that encoder-free Mel inputs preserve low-level acoustic cues that are beneficial for paralinguistic understanding.
    \item We show that multimodal pre-training (Phi-4-MM) initialization is critical when training data is limited and provide layer-freezing evidence that lower/middle LLM layers contribute most to speech adaptation.
    \item We include a preliminary TTS proof-of-concept, highlighting the remaining difficulty of high-fidelity continuous Mel generation with an LLM backbone.
\end{itemize}

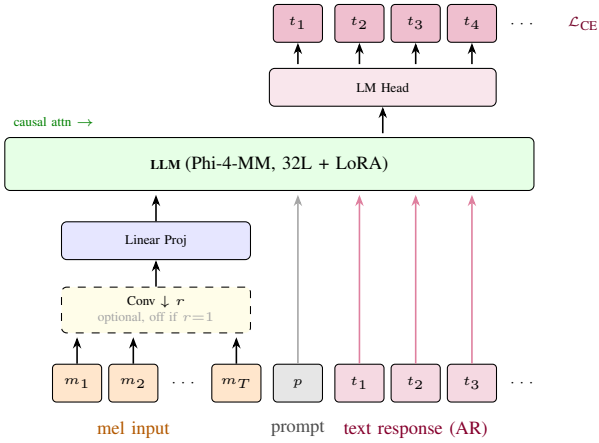
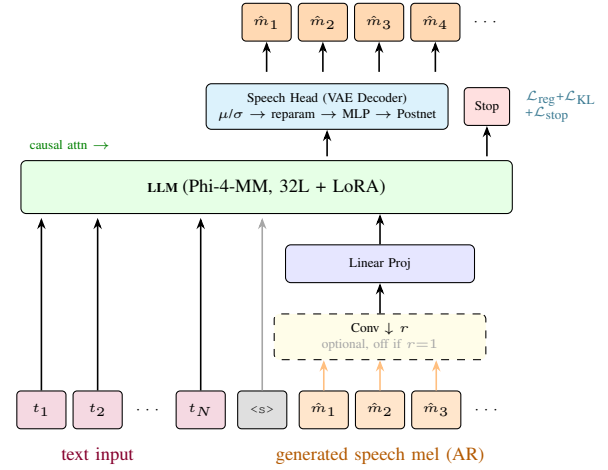
\begin{figure*}[t]
    \centering
    \begin{subfigure}[t]{0.48\textwidth}
        \centering
\begin{tikzpicture}[
    node distance=0.35cm,
    tok/.style={rectangle, draw, rounded corners=1.5pt, minimum width=0.65cm, minimum height=0.5cm, align=center, font=\tiny, inner sep=1pt},
    wide/.style={rectangle, draw, rounded corners=2pt, minimum height=0.5cm, align=center, font=\tiny},
    arrow/.style={-{Stealth[length=1.5mm]}, semithick},
    darrow/.style={-{Stealth[length=1.5mm]}, semithick, dashed, gray!70},
]

\node[tok, fill=orange!20] (m1) {$m_1$};
\node[tok, fill=orange!20, right=0.08cm of m1] (m2) {$m_2$};
\node[font=\tiny, right=0.04cm of m2] (md) {$\cdots$};
\node[tok, fill=orange!20, right=0.04cm of md] (mT) {$m_T$};

\node[tok, fill=gray!20, right=0.15cm of mT] (p1) {$p$};

\node[tok, fill=purple!15, right=0.15cm of p1] (t1) {$t_1$};
\node[tok, fill=purple!15, right=0.08cm of t1] (t2) {$t_2$};
\node[tok, fill=purple!15, right=0.08cm of t2] (t3) {$t_3$};
\node[font=\tiny, right=0.04cm of t3] (td) {$\cdots$};

\node[font=\scriptsize, below=0.12cm of m2, text=orange!70!black] {mel input};
\node[font=\scriptsize, below=0.12cm of p1, text=gray!60!black] {prompt};
\node[font=\scriptsize, below=0.12cm of t2, text=purple!70!black] {text response (AR)};

\node[wide, fill=yellow!10, dashed, above=0.4cm of m2, minimum width=2.5cm, xshift=0.3cm] (conv) {Conv $\downarrow r$\\{\tiny\color{gray!70} optional, off if $r{=}1$}};

\node[wide, fill=blue!10, above=0.35cm of conv, minimum width=2.5cm] (proj) {Linear Proj};

\node[wide, fill=green!10, above=0.4cm of proj, minimum width=7cm, minimum height=0.7cm, xshift=1.5cm] (llm) {\textbf{LLM} {\scriptsize (Phi-4-MM, 32L + LoRA)}};

\node[wide, fill=purple!10, above=0.4cm of llm, minimum width=3.0cm, xshift=1.5cm] (lmhead) {LM Head};

\node[tok, fill=purple!25] at ([yshift=4.5cm]p1.north) (o1) {$t_1$};
\node[tok, fill=purple!25] at ([yshift=4.5cm]t1.north) (o2) {$t_2$};
\node[tok, fill=purple!25] at ([yshift=4.5cm]t2.north) (o3) {$t_3$};
\node[tok, fill=purple!25] at ([yshift=4.5cm]t3.north) (o4) {$t_4$};
\node[font=\tiny, right=0.04cm of o4] (od) {$\cdots$};

\draw[arrow] (m1.north) -- ([yshift=-0.05cm]conv.south -| m1.north);
\draw[arrow] (m2.north) -- ([yshift=-0.05cm]conv.south -| m2.north);
\draw[arrow] (mT.north) -- ([yshift=-0.05cm]conv.south -| mT.north);

\draw[arrow] (conv.north) -- (proj.south);

\draw[arrow] (proj.north) -- ([yshift=-0.05cm]llm.south -| proj.north);

\draw[arrow, gray!70] (p1.north) -- ([yshift=-0.05cm]llm.south -| p1.north);

\draw[arrow, purple!50] (t1.north) -- ([yshift=-0.05cm]llm.south -| t1.north);
\draw[arrow, purple!50] (t2.north) -- ([yshift=-0.05cm]llm.south -| t2.north);
\draw[arrow, purple!50] (t3.north) -- ([yshift=-0.05cm]llm.south -| t3.north);

\draw[arrow] ([yshift=0.05cm]llm.north -| lmhead.south) -- (lmhead.south);

\draw[arrow] ([yshift=0.05cm]lmhead.north -| o1.south) -- (o1.south);
\draw[arrow] ([yshift=0.05cm]lmhead.north -| o2.south) -- (o2.south);
\draw[arrow] ([yshift=0.05cm]lmhead.north -| o3.south) -- (o3.south);
\draw[arrow] ([yshift=0.05cm]lmhead.north -| o4.south) -- (o4.south);

\node[font=\tiny, text=green!50!black, above=0.02cm of llm.north west, anchor=south west] {causal attn $\rightarrow$};

\node[font=\tiny, text=purple!60!black, right=0.15cm of od] {$\mathcal{L}_{\text{CE}}$};

\end{tikzpicture}
        \caption{Speech understanding: Mel spectrogram is passed through convolution layers if downsampling is required, and linearly projected into the LLM embedding space. The LLM autoregressively generates text responses for ASR and speech understanding tasks.}
        \label{fig:arch_asr}
    \end{subfigure}
    \hfill
    \begin{subfigure}[t]{0.48\textwidth}
        \centering
\begin{tikzpicture}[
    node distance=0.35cm,
    tok/.style={rectangle, draw, rounded corners=1.5pt, minimum width=0.65cm, minimum height=0.5cm, align=center, font=\tiny, inner sep=1pt},
    wide/.style={rectangle, draw, rounded corners=2pt, minimum height=0.5cm, align=center, font=\tiny},
    arrow/.style={-{Stealth[length=1.5mm]}, semithick},
    darrow/.style={-{Stealth[length=1.5mm]}, semithick, dashed, gray!70},
]

\node[tok, fill=purple!15] (t1) {$t_1$};
\node[tok, fill=purple!15, right=0.08cm of t1] (t2) {$t_2$};
\node[font=\tiny, right=0.04cm of t2] (td) {$\cdots$};
\node[tok, fill=purple!15, right=0.04cm of td] (tN) {$t_N$};

\node[tok, fill=gray!25, right=0.15cm of tN] (bos) {\texttt{<s>}};

\node[tok, fill=orange!20, right=0.15cm of bos] (m1) {$\hat{m}_1$};
\node[tok, fill=orange!20, right=0.08cm of m1] (m2) {$\hat{m}_2$};
\node[tok, fill=orange!20, right=0.08cm of m2] (m3) {$\hat{m}_3$};
\node[font=\tiny, right=0.04cm of m3] (md) {$\cdots$};

\node[font=\scriptsize, below=0.12cm of t2, text=purple!70!black] {text input};
\node[font=\scriptsize, below=0.12cm of m2, text=orange!70!black] {generated speech mel (AR)};

\node[wide, fill=yellow!10, dashed, above=0.4cm of m2, minimum width=2.8cm] (conv) {Conv $\downarrow r$\\{\tiny\color{gray!70} optional, off if $r{=}1$}};

\node[wide, fill=blue!10, above=0.4cm of conv, minimum width=2.5cm] (proj) {Linear Proj};

\node[wide, fill=green!10, above=0.4cm of proj, minimum width=6.5cm, minimum height=0.7cm, xshift=-1.5cm] (llm) {\textbf{LLM} {\scriptsize (Phi-4-MM, 32L + LoRA)}};

\node[wide, fill=cyan!12, above=0.4cm of llm, minimum width=3.2cm, xshift=0.8cm] (vae) {Speech Head (VAE Decoder)\\{\tiny $\mu$/$\sigma$ $\to$ reparam $\to$ MLP $\to$ Postnet}};

\node[tok, fill=red!12, right=0.2cm of vae, minimum width=0.6cm] (stop) {Stop};

\node[tok, fill=orange!30, above=0.55cm of vae, xshift=-0.8cm] (o1) {$\hat{m}_1$};
\node[tok, fill=orange!30, right=0.08cm of o1] (o2) {$\hat{m}_2$};
\node[tok, fill=orange!30, right=0.08cm of o2] (o3) {$\hat{m}_3$};
\node[tok, fill=orange!30, right=0.08cm of o3] (o4) {$\hat{m}_4$};
\node[font=\tiny, right=0.04cm of o4] (od) {$\cdots$};

\draw[arrow] (t1.north) -- ([yshift=-0.05cm]llm.south -| t1.north);
\draw[arrow] (t2.north) -- ([yshift=-0.05cm]llm.south -| t2.north);
\draw[arrow] (tN.north) -- ([yshift=-0.05cm]llm.south -| tN.north);

\draw[arrow, gray!70] (bos.north) -- ([yshift=-0.05cm]llm.south -| bos.north);

\draw[arrow, orange!50] (m1.north) -- (m1.north |- conv.south);
\draw[arrow, orange!50] (m2.north) -- (m2.north |- conv.south);
\draw[arrow, orange!50] (m3.north) -- (m3.north |- conv.south);
\draw[arrow] (conv.north) -- (proj.south);
\draw[arrow] (proj.north) -- (proj.north |- llm.south);

\draw[arrow] ([yshift=0.05cm]llm.north -| vae.south) -- (vae.south);

\draw[arrow] ([yshift=0.05cm]llm.north -| stop.south) -- (stop.south);

\draw[arrow] ([yshift=0.15cm]vae.north -| o1.south) -- (o1.south);
\draw[arrow] ([yshift=0.15cm]vae.north -| o2.south) -- (o2.south);
\draw[arrow] ([yshift=0.15cm]vae.north -| o3.south) -- (o3.south);
\draw[arrow] ([yshift=0.15cm]vae.north -| o4.south) -- (o4.south);

\node[font=\tiny, text=green!50!black, above=0.02cm of llm.north west, anchor=south west] {causal attn $\rightarrow$};

\node[font=\tiny, text=cyan!50!black, right=0.08cm of stop.east, align=left] {$\mathcal{L}_{\text{reg}}$+$\mathcal{L}_{\text{KL}}$\\+$\mathcal{L}_{\text{stop}}$};

\end{tikzpicture}
        \caption{Speech generation: Given text input, speech Mel chunks are autoregressively generated by the LLM and a VAE decoder (linear $\to$ $\mu$/$\sigma$, reparameterize, residual MLP, postnet) similarly to MELLE~\cite{meng2025melle}. TTS is the representative generation task in this work.}
        \label{fig:arch_tts}
    \end{subfigure}
    \vspace{-0.05cm}
    \caption{Architecture of Mel-LLM for (a) speech understanding and (b) exploratory speech generation. Both directions share the same LLM backbone with LoRA adaptation. The encoder-free understanding path removes the dedicated speech encoder blocks, while retaining linear projection and lightweight convolutional downsampling when required.}
    \label{fig:architecture}
    \vspace{-0.08cm}
\end{figure*}

\vspace{-0.05cm}
\section{Related Work}

Recent works~\cite{FathullahWLJSLG24,ShiJXXZWSZY24,abouelenin2025phi,tang2024salmonn,fan2024alignformer,wu2025jointdecoding,hu2025slms2st,deng2026cotasr} have established the encoder-projector-LLM paradigm for speech understanding. These systems typically use large pre-trained speech encoders (Whisper~\cite{radford2023robust}, HuBERT~\cite{hsu2021hubert}) to extract semantic features before feeding them to the LLM.

\textbf{Encoder-free multimodal models.} In vision, Fuyu~\cite{fuyu2023} first demonstrated that raw image patches can be directly projected into an LLM without a vision encoder. ELVA~\cite{li2025breaking} extends this direction to encoder-free video-language understanding with hierarchical token merging and video guidance. These works focus on multimodal understanding rather than generation, complementing our speech-understanding experiments~\cite{diao2024unveiling}. More recently, Tuna-2~\cite{liu2026tuna2} achieves state-of-the-art results for both multimodal understanding and generation, with simple patch embedding layers and no modular encoder. Representation Forcing~\cite{wang2026repforcing} further eliminates the VAE bottleneck by forcing the decoder to predict visual representations as intermediate tokens before pixel generation. An and colleagues~\cite{an2026nativemm} provide a comprehensive roadmap toward native multimodal modeling (NMM), distinguishing early-fusion architectures from non-native paradigms. These vision advances motivate our encoder-free exploration in speech.

\textbf{Continuous mel-spectrogram TTS.} MELLE~\cite{meng2025melle} pioneered autoregressive continuous mel generation without vector quantization, using regression loss with spectrogram flux and variational inference. MELA-TTS~\cite{an2025melatts} extends this with a joint Transformer-diffusion framework and representation alignment from a pre-trained ASR encoder. MELD~\cite{yeh2026meld} introduces discrete latent variables on Mel spectrogram with joint encoder-LM optimization for both TTS and STT. WavFlow~\cite{zhou2026wavflow} pushes further by generating audio directly in raw waveform space without intermediate representations. We use TTS as an exploratory generation setting; the main contribution of this paper is encoder-free speech understanding.

\textbf{Speech-text unified models.} SpeechGPT~\cite{zhang2023speechgpt}, VoxtLM~\cite{maiti2024voxtlm}, and Spirit LM~\cite{nguyen2024spiritlm} explore unified speech-text generation, typically using discrete speech tokens. dMel~\cite{bai2024dmel} simplifies this direction by discretizing Mel-filterbank channels into intensity-bin tokens, enabling speech modeling with language-model-style discrete units. Our work differs by operating directly on continuous Mel spectrogram features as LLM inputs, avoiding discrete tokenization.

To our knowledge, this is the first academic publication providing detailed encoder-based vs.\ encoder-free comparisons for Speech-LLMs across data scales.\footnote{During the preparation of this manuscript, Google announced Gemma~4~12B (Jun.\ 2026) which removes audio encoders and later reports in ~\cite{team2026gemma}.}
\section{Method}
\label{sec:method}
\vspace{-0.05cm}

\subsection{Architecture Overview}

The Mel-LLM architecture is illustrated in Figure~\ref{fig:architecture}. We build upon the standard Speech-LLM framework~\cite{FathullahWLJSLG24,ShiJXXZWSZY24} but systematically simplify the speech encoder component. Our system supports both speech understanding, e.g. ASR, and speech generation (TTS in this work), using Phi-4-MultiModal (Phi-4-MM)~\cite{abouelenin2025phi} as the base model. The overall architecture is simple, sharing most of architecture knowledge from LLM, especially the Phi-4-MM. Please refer to ~\cite{abouelenin2025phi} for more details.

\subsection{Speech Understanding Input Path}

Given a speech signal, we first extract 80-dimensional log-Mel spectrogram features. These features are then processed through the following pipeline:

\textbf{MVN Normalization.} We apply  mean and variance normalization (MVN) on  using pre-computed statistics from the training set.

\textbf{Lightweight Convolution layers (Optional).} The normalized spectrogram of shape $[T, 80]$ is projected into a hidden space with a reduced sequence length at $T/r$ where $r$ is the time reduction factor (we use $r=2$ for TTS and $8$ for ASR). Ablation study on $r$ is conducted in Section~\ref{ssec:asr_ablation}, and the convolution layers are disabled for the $r=1$ setting.



\textbf{Linear Projection.} A single linear layer projects the features into the LLM's hidden dimension:
\begin{align}
    E^s = W_{\text{proj}} \cdot e + b_{\text{proj}}, \quad W_{\text{proj}} \in \mathbb{R}^{d_{\text{LLM}} \times d_{\text{enc}}}
\end{align}
In the encoder-free setting, this projection is randomly initialized (as opposed to being pre-trained alongside an encoder), and the LLM learns to interpret raw spectral features through its own parameters.

\textbf{LLM Processing.} The projected speech embeddings $E^s$ are concatenated with text prompt embeddings $E^p$ and fed into the LLM, which autoregressively generates the transcription:
\begin{align}
    \hat{T} = \text{LLM}(\text{Concat}(E^s, E^p))
\end{align}
The LLM is adapted using LoRA~\cite{HuSWALWWC22} with rank 320.


\subsection{Exploratory Speech Generation Path}

We also test whether the same LLM backbone can autoregressively generate continuous Mel spectrograms for TTS. Given text input, the LLM predicts hidden states at speech positions, and a lightweight Mel head decodes each state into Mel frames using a VAE-style projection, residual MLP, postnet, and stop predictor, following the continuous-generation setup of MELLE~\cite{meng2025melle}. The TTS objective combines reconstruction, KL, stop-token, and temporal smoothness losses. This component is intended as a proof-of-concept for future unified modeling rather than a competitive TTS system.

\vspace{-0.05cm}
\section{Experimental Settings}
\label{sec:exp_set}

\subsection{Model Configuration}
Our model is built upon Phi-4-MM~\cite{abouelenin2025phi}. The LLM has a hidden dimension of 3072, 32 layers, 24 attention heads, and 8 KV heads. We use LoRA with $r=320$, $\alpha=640$ for linear layers in attention and MLP blocks. For ASR, the main Conformer encoder blocks are removed while NeMoConv layers are preserved for downsampling purposes. For TTS, the $r$ for mel input is set to 2 to align with the best setting in MELLE. The VAE decoder uses 3 residual MLP layers with dimension 256, followed by a 5-layer Conv1D postnet. Table~\ref{tab:model_config} summarizes the main configuration of the 12.5Hz Mel-LLM setting. We report measured speedup rather than theoretical FLOPs because the effective compute depends strongly on sequence length, batching, and fused attention kernels.

\begin{table}[!t]
\centering
\caption{Model configuration for the main 12.5Hz Mel-LLM setting. The model is encoder-free in the sense that dedicated Transformer/Conformer speech encoder blocks are removed; only lightweight convolutional downsampling is retained.}
\vspace{-0.05cm}
\resizebox{\columnwidth}{!}{
\begin{tabular}{l|l}
\toprule
Item & Configuration \\
\midrule
Base LLM & Phi-4-MM, 32 layers, $d{=}3072$ \\
LoRA adaptation & $r{=}320$, $\alpha{=}640$ on attention/MLP projections \\
LoRA trainable params & $\sim$503M \\
Frozen params & LLM layers, norm, embeddings, and LM head \\
Removed frontend & Transformer/Conformer speech encoder blocks \\
Retained frontend & NeMoConv downsampling, $8\times$, 1024 channels \\
Frontend params & 12.6M \\
Audio projection & 2-layer MLP, 1024$\to$3072$\to$3072, 12.6M \\
Total non-LLM audio stack & 25.2M params \\
Main token rate & 12.5Hz \\
Measured training speed & 1.57$\times$ vs. encoder-based baseline \\
\bottomrule
\end{tabular}}
\label{tab:model_config}
\vspace{-0.05cm}
\end{table}

\begin{table*}[!t]
\centering
\caption{ASR performance (WER\%) on OpenASR Leaderboard. ``Encoder'' and ``LoRA'' columns indicate initialization: Pretrained (from Phi-4-MM), Random, or None. Mel-LLM removes the dedicated Transformer/Conformer speech encoder blocks while retaining lightweight convolutional downsampling.}
\label{tab:openasr}
\vspace{-0.05cm}
\resizebox{\textwidth}{!}{
\begin{tabular}{l|cc|cccccccc|c}
\toprule
System & Encoder & LoRA & AMI & Earnings22 & Gigaspeech & LS-clean & LS-other & SPGISpeech & TED-LIUM & VoxPopuli & \textbf{Avg} \\
\midrule
Whisper-Large-V3~\cite{radford2023robust} & N/A & N/A & 15.95 & 11.29 & 10.02 & 2.01 & 3.91 & 2.94 & 3.86 & 9.54 & 7.44 \\
Gemma-4-12B-it & None & N/A & 131.52 & 29.17 & 19.66 & 3.83 & 9.26 & 6.78 & 6.19 & 7.68 & 26.76 \\
\midrule
Phi-4-MM & N/A & N/A & 11.69 & 10.16 & 9.78 & 1.68 & 3.83 & 3.13 & 2.90 & 5.91 & 6.14 \\
\quad + FT & Pretrained & Pretrained & 11.16 & 9.57 & 9.45 & 1.32 & 2.95 & 1.70 & 2.70 & 6.03 & 5.61 \\
\quad + Random Enc FT & Random & Pretrained & 12.19 & 14.31 & 10.38 & 1.62 & 4.27 & 2.04 & 3.29 & 7.65 & 6.97 \\
\midrule
Mel-LLM (Phi-4-MM init) & None & Pretrained & 12.91 & 12.99 & 10.95 & 1.70 & 4.83 & 2.28 & 3.55 & 7.76 & \textbf{7.12} \\
Mel-LLM (Random init) & None & Random & 13.65 & 11.98 & 11.38 & 1.83 & 5.50 & 2.47 & 4.42 & 8.25 & 7.44 \\
\bottomrule
\end{tabular}}
\par\vskip 0.8ex
\noindent\parbox{\textwidth}{\footnotesize\emph{Note:} The OpenASR leaderboard announced on 2026-05-20 that TED-LIUM was removed from the leaderboard suite. The OpenASR numbers in this table were obtained before this change and follow the earlier benchmark definition that included TED-LIUM.}
\vspace{-0.05cm}
\end{table*}

\begin{table}[!t]
\centering
\caption{Performance gap (WER\%) between encoder-initialized and encoder-free models at different data scales on in-house test sets. With limited public data, encoder-free degrades notably. With 10$\times$ scaled anonymized in-house data, the gap narrows significantly, confirming data scaling as the key enabler.}
\vspace{-0.05cm}
\resizebox{\columnwidth}{!}{
\begin{tabular}{l|c|cc|cc}
\toprule
Test Set & Enc-Init & \makecell{Encoder-Free\\(limited data)} & $\Delta$ rel. & \makecell{Encoder-Free\\(10$\times$ scaled)} & $\Delta$ rel. \\
\midrule
Call Center & 15.92 & 18.28 & +14.8\% & 16.74 & +5.2\% \\
Conversation & 15.83 & 17.10 & +8.0\% & 16.25 & +2.7\% \\
Dictation & 5.80 & 6.40 & +10.3\% & 5.99 & +3.3\% \\
\midrule
Average & 12.52 & 13.93 & +11.3\% & 12.99 & +3.8\% \\
\bottomrule
\end{tabular}}
\label{tab:scaling}
\vspace{-0.05cm}
\end{table}

\begin{table*}[!t]
\centering
\caption{Ablation on token rate. Lower token rates reduce sequence length and improve training speed but degrade quality. Mel-LLM at 12.5Hz achieves 1.57$\times$ speedup over the encoder-based baseline while maintaining competitive WER.}
\vspace{-0.05cm}
\resizebox{\textwidth}{!}{
\begin{tabular}{l|c|cccccccc|c|c}
\toprule
System & Token Rate & AMI & Earnings22 & Gigaspeech & LS-clean & LS-other & SPGISpeech & TED-LIUM & VoxPopuli & \textbf{Avg} & Speedup \\
\midrule
Phi-4-MM-FT-Base & 12.5Hz & 12.19 & 14.31 & 10.38 & 1.62 & 4.27 & 2.04 & 3.29 & 7.65 & 6.97 & 1.0$\times$ \\
\midrule
\multirow{5}{*}{Mel-LLM} & 100Hz & 12.34 & 10.56 & 10.41 & 1.63 & 4.50 & 2.20 & 3.29 & 7.74 & 6.58 & 0.33$\times$ \\
 & 50Hz & 12.65 & 10.89 & 10.67 & 1.64 & 4.59 & 2.14 & 3.24 & 7.84 & 6.71 & 0.65$\times$ \\
 & 25Hz & 13.18 & 13.69 & 10.80 & 1.73 & 4.77 & 2.15 & 3.38 & 7.96 & 7.21 & 1.09$\times$ \\
 & 12.5Hz & 12.91 & 12.99 & 10.95 & 1.70 & 4.83 & 2.28 & 3.55 & 7.76 & \textbf{7.12} & \textbf{1.57$\times$} \\
 & 6.25Hz & 14.80 & 15.13 & 11.82 & 1.86 & 5.70 & 2.49 & 3.91 & 8.43 & 8.02 & 1.88$\times$ \\
\bottomrule
\end{tabular}}
\label{tab:downsampling}
\vspace{-0.05cm}
\end{table*}

\subsection{Training Data and Settings}

For ASR experiments on the OpenASR leaderboard, we train exclusively on publicly available English speech corpora, totaling approximately 31M utterances ($\sim$64k hours). The training set comprises: LibriSpeech~\cite{panayotov2015librispeech} (960h), GigaSpeech~\cite{chen2021gigaspeech} (10kh), Multilingual LibriSpeech (MLS) English~\cite{pratap2020mls} (44kh), SPGISpeech~\cite{oneill2021spgispeech} (5kh), CommonVoice 15 English~\cite{ardila2020commonvoice}, VoxPopuli English~\cite{wang2021voxpopuli}, TED-LIUM~\cite{hernandez2018tedlium3}, AMI~\cite{carletta2005ami}, Earnings-22~\cite{rio2022earnings22}, and FLEURS English~\cite{conneau2023fleurs}. No proprietary or internal data is used for these experiments. 

For the expanded speech understanding experiments, we augment the ASR mixture with English speech QA and audio/speech understanding data. The final mixture contains 55.3M weighted examples, which agrees with the Phi-4-MM post-training data but a subset. The speech-QA portion includes multi-turn SQA data, WavLLM-style QA data following WavLLM~\cite{hu2024wavllm}. The speech/audio understanding portion follows LTU-AS~\cite{gong2023joint}, including audio/music instruction data, VoxCeleb speaker data, MOSEI emotion data, AudioSet/FSD50K/AudioCaps/Clotho-style data. These sources intentionally cover task families and label spaces related to the evaluation suite, but we follow the train/evaluation splits used by the corresponding WavLLM and LTU-AS protocols and do not include the held-out evaluation examples in training. These sources have different duration distributions, and we use utterance-based dynamic batching: larger batches for short ASR chunks and smaller batches for long SQA or audio understanding examples.

We train using DeepSpeed ZeRO Stage-1 on 16 NVIDIA H100 GPUs. We use the AdamW optimizer with peak learning rate $1 \times 10^{-4}$, linear warmup-decay scheduling (9000 warmup steps), gradient clipping at 1.0, and effective batch size of 512. The base LLM layers are frozen with only LoRA trainable. The training data is iterated for three times and we do not find additional gains with more sweeps. We ablate the time reduction rate for performance and speed trade-off, the layer-wise LoRA initialization and training behaviors for speech encoding insights. For the production-scale experiments in Table~\ref{tab:scaling}, we use about 10$\times$ anonymized in-house training data for one iteration with a similar setup.

For TTS experiments, we use Libriheavy 50k hours English data for preliminary training. We use dropout of 0.5 for both input linear projection layers and the output Mel head. The KL loss weight is 0.05, stop head loss weight is 1.0 and flux loss is 0.5 respectively. The libriheavy data is swept for 5 epochs.

\subsection{Evaluation}

For ASR, we evaluate on the OpenASR Leaderboard~\cite{openasr}, which includes diverse benchmarks covering various domains and acoustic conditions, reported in word error rate (WER). We additionally report production-level scaling experiments on anonymized in-house test sets covering call center, conversation, and dictation scenarios, also reported in WER.

For speech understanding, we evaluate representative non-ASR and spoken-QA abilities: ESC-50 audio classification, IEMOCAP emotion classification, gender classification, VoxCeleb age prediction, GTZAN music genre classification, in LTU~\cite{gong2023joint}, WavLLM speaker understanding~\cite{hu2024wavllm}, MMAU-mini~\cite{sakshi2024mmau}, and MMLU speech English~\cite{hendrycks2021measuring}. These tasks are intentionally heterogeneous: some depend heavily on acoustic or paralinguistic cues, while others require semantic grounding and retained text-side knowledge. Mel-LLM uses one speech-understanding checkpoint trained with the mixture in Section~\ref{sec:exp_set}; Gemma-4-12B-it is evaluated directly with no task-specific tuning. All systems use the benchmark-provided task prompts and greedy decoding. Classification tasks are scored by normalized label or choice matching: ESC-50, GTZAN, and MMAU use choice/text matching; IEMOCAP, gender, and speaker verification use label-keyword normalization; MMLU-speech extracts the final A/B/C/D answer from both canonical ``Answer: X'' and verbose answer forms. Invalid or unmatched classification outputs are counted as wrong. For age regression, we extract the first numeric age prediction or range midpoint; non-numeric outputs are reported as invalid and excluded.

For TTS, we evaluate zero-shot synthesis on LibriSpeech test-clean using WER (measured by Whisper-large-v3~\cite{radford2023robust} for intelligibility) and UTMOS~\cite{saeki2022utmos} for perceptual quality.

\vspace{-0.05cm}
\section{Experimental Results}

\begin{table*}[!t]
\centering
\caption{Ablation on layer-wise initialization and freezing upper LoRA layers (initialized from Phi-4-MM). ``Freeze L$k$--31'' keeps LoRA layers $k$ through 31 at Phi-4-MM initialization without fine-tuning. Results suggest that layers 24+ contribute minimally, consistent with these upper layers already focusing on high-level semantics in Phi-4-MM.}
\vspace{-0.05cm}
\begin{tabular}{l|cccccccc|c}
\toprule
System & AMI & Earnings22 & Gigaspeech & LS-c & LS-o & SPGISpeech & TED-LIUM & VoxPopuli & \textbf{Avg} \\
\midrule
Phi-4-MM-FT-Base & 12.19 & 14.31 & 10.38 & 1.62 & 4.27 & 2.04 & 3.29 & 7.65 & 6.97 \\
\midrule
Mel-LLM (Random init) & 13.65 & 11.98 & 11.38 & 1.83 & 5.50 & 2.47 & 4.42 & 8.25 & 7.44 \\
\midrule
Mel-LLM (all LoRA) & 12.91 & 12.99 & 10.95 & 1.70 & 4.83 & 2.28 & 3.55 & 7.76 & 7.12 \\
\quad + init. and freeze L16--31 & 14.08 & 13.43 & 11.78 & 2.25 & 6.67 & 2.86 & 3.96 & 8.48 & 7.94 \\
\quad + init. and freeze L20--31 & 13.72 & 14.34 & 11.38 & 1.96 & 6.13 & 2.64 & 3.89 & 8.08 & 7.77 \\
\quad + init. and freeze L24--31 & 13.76 & 12.72 & 11.26 & 1.95 & 5.54 & 2.46 & 3.70 & 8.05 & 7.43 \\
\quad + init. and freeze L28--31 & 13.66 & 12.76 & 11.19 & 1.85 & 5.56 & 2.43 & 3.70 & 8.07 & 7.40 \\

\bottomrule
\end{tabular}
\label{tab:layer_ablation}
\vspace{-0.05cm}
\end{table*}

\begin{table*}[!t]
\centering
\caption{Speech/audio understanding and spoken QA results. Higher ACC and lower MAE are better. Mel-LLM improves tasks that rely on acoustic/paralinguistic cues, while encoder-anchored Phi-4-MM and Gemma-4-12B-it remain stronger on knowledge-intensive MMLU-speech.}
\vspace{-0.05cm}
\resizebox{\textwidth}{!}{
\begin{tabular}{l|ccccc|cccc}
\toprule
System & ESC-50 & IEMOCAP & Gender & Age & GTZAN & WavLLM-spk. & MMAU-mini & MMLU-speech & MMLU-Text \\
 & ACC $\uparrow$ & ACC $\uparrow$ & ACC $\uparrow$ & MAE $\downarrow$ & ACC $\uparrow$ & ACC $\uparrow$ & ACC $\uparrow$ & ACC $\uparrow$ & ACC $\uparrow$ \\
\midrule
LTU-AS~\cite{gong2023joint} & \textbf{80.80} & 65.20 & 90.80 & \textbf{7.30} & 50.30 & -- & -- & -- & -- \\
Gemma-4-12B-it & 6.15 & 13.62 & 71.68 & 126.68 & 12.15 & 56.53 & 31.60 & \textbf{57.85} & \textbf{75.18} \\
Phi-4-MM & 35.95 & 40.69 & 92.57 & 8.06 & 40.54 & 76.64 & 50.70 & 53.12 & 61 \\
\midrule
Mel-LLM & 67.85 & 66.16 & 97.73 & 8.39 & 46.85 & 87.98 & 54.10 & 41.30 & 58.68 \\
Mel-LLM + ASR init. & 67.00 & \textbf{72.36} & \textbf{98.21} & 7.56 & \textbf{53.55} & \textbf{92.10} & \textbf{56.30} & 36.47 & 56.73 \\
\bottomrule
\end{tabular}}
\label{tab:audio_understanding}
\vspace{-0.05cm}
\end{table*}

\subsection{ASR: Main Results and Scaling}

We present the ASR results of Mel-LLM in Table~\ref{tab:openasr}. The conventional attention-based encoder-decoder baseline, Whisper-large-v3, and Gemma-4-12B-it results are included. Gemma-4-12B-it is unstable for noisy long-form ASR, with frequent refusals or responses indicating that speech is not detected on AMI-like inputs.

\textbf{Phi-4-MM initialization is critical at limited data scale.} The Phi-4-MM initialization indicates using LoRA parameters only while convolution frontend and adapters are randomly initialized. The encoder-free Mel-LLM with Phi-4-MM initialization achieves 7.12\% average WER on OpenASR (Table~\ref{tab:openasr}), only 0.15\% behind the random-encoder baseline (6.97\%) that still uses a trainable encoder. Random initialization of the LLM degrades to 7.44\%, a further 0.32\% drop. This confirms that multimodal pre-training provides useful inductive biases for interpreting raw spectral features when training data is limited, even though the encoder-based and encoder-free variants expose the LLM to different hidden-space distributions. In other words, the pretrained LoRA already contains speech-text alignment knowledge that a randomly initialized model must learn from scratch.

\textbf{Data scaling closes the encoder-free gap.} Table~\ref{tab:scaling} reveals that the performance gap between encoder-initialized and encoder-free models shrinks dramatically with data scale. With limited public data, the encoder-free model shows notable degradation across all in-house test sets (12.52\%$\to$13.93\% average WER). However, with 10$\times$ scaled anonymized in-house data, this gap narrows to only +3.8\% relative (12.52\%$\to$12.99\%). This demonstrates that sufficient data is the key for encoder-free solutions---given enough training signal, the LLM learns to perform implicit speech encoding without relying on a dedicated encoder or initialization.

\subsection{ASR: Ablation Studies}
\label{ssec:asr_ablation}

\textbf{Downsampling rate vs.\ training efficiency.} Table~\ref{tab:downsampling} shows the trade-off between token rate and ASR quality. Phi-4-MM-FT-Base uses 8$\times$ downsampling, corresponding to a 12.5Hz token rate. Higher token rates (100Hz, 50Hz) yield lower WER (6.58\%, 6.71\%) but are significantly slower due to longer sequence lengths. Notably, the 25Hz and 12.5Hz settings achieve similar performance (7.21\% vs. 7.12\%). We attribute this to the Phi-4-MM LoRA initialization: the pretrained LoRA was trained with a 12.5Hz token rate in Phi-4-MM, so Mel-LLM at 12.5Hz benefits directly from this initialization, while 25Hz must adapt the pretrained weights to a mismatched frame rate. The 12.5Hz setting achieves the best quality-speed trade-off with 1.57$\times$ training speedup over the encoder-based baseline. This speedup comes from removing the main encoder blocks while keeping lightweight convolutional downsampling, making Mel-LLM beneficial for both architectural simplicity and computational efficiency.

\textbf{Ablation on layer-wise initialization in LLM.} Table~\ref{tab:layer_ablation} investigates which LLM layers are important for speech adaptation by freezing upper LoRA layers at their Phi-4-MM initialization. Freezing L28--31 causes only 0.28\% degradation (7.12\%$\to$7.40\%), and freezing L24--31 yields 7.43\%---nearly identical to the fully trainable setting. However, extending the freeze to L20--31 (7.77\%) and L16--31 (7.94\%) leads to progressively larger drops. This pattern suggests that layers 24+ in the pretrained Phi-4-MM already encode high-level language semantics---reasoning, discourse structure, and text generation capabilities---that transfer well to ASR without speech-specific adaptation. In contrast, layers 0--23 appear more important for adapting raw spectral patterns into linguistically meaningful representations. This is consistent with findings that lower Transformer layers capture local features while upper layers handle abstract semantics. The frozen upper layers can still perform text generation effectively because their pretrained weights already handle the ``language modeling'' part of ASR; what they lack is the low-level acoustic adaptation that the lower layers must learn.

\subsection{Understanding: Acoustic Cues vs. Semantic Anchoring}

Table~\ref{tab:audio_understanding} extends the evaluation from ASR to broader speech/audio understanding. The most consistent gains appear on tasks where the answer is carried by acoustic structure rather than lexical content. Compared with Phi-4-MM, Mel-LLM improves IEMOCAP emotion classification (40.69$\to$66.16), gender classification (92.57$\to$97.73), WavLLM speaker understanding (76.64$\to$87.98), and MMAU-mini (50.70$\to$54.10). With ASR initialization, the same trend becomes stronger: emotion reaches 72.36, gender 98.21, WavLLM speaker 92.10, and GTZAN music genre 53.55. These results support the central hypothesis of this paper from a new angle: when the dedicated Transformer/Conformer speech encoder is removed, the LLM is forced to directly model information that the semantic speech encoder in Phi-4-MM may compress away, including speaker traits, prosody, timbre, emotion, and music texture. Since the expanded Mel-LLM data recipe follows the speech/audio instruction data used for the encoder-based Phi-4-MM setting, these gains suggest that direct Mel input is beneficial for paralinguistic cues. We also evaluate the open-source Gemma-4-12B-it model (\texttt{google/gemma-4-12B-it}), which accepts native audio input. Its low paralinguistic scores should be interpreted with output-format behavior: after task-specific answer normalization, invalid rates are high on IEMOCAP emotion (71.07\%) and age prediction (96.81\%; only 73/2288 numeric predictions), moderate on MMAU-mini (29.00\%) and MMLU-speech (14.89\%), and low on gender, WavLLM speaker, ESC-50, and GTZAN ($<9\%$), indicating that native waveform input alone does not guarantee robust acoustic reasoning.

The MMLU-speech result shows the opposite side of this trade-off. Phi-4-MM remains stronger on knowledge-intensive spoken QA, while Mel-LLM lags behind. The text-only MMLU column helps diagnose this drop: Mel-LLM remains close to Phi-4-MM on text-only MMLU (61.0$\to$58.68), and the ASR-initialized variant is only slightly lower (56.73). Thus, under our current evidence, the MMLU-speech degradation is unlikely to be caused solely by catastrophic loss of text reasoning ability. Instead, the gap between text and speech QA is much larger for the encoder-free models: Gemma-4-12B-it drops from 75.18 on text MMLU to 57.85 on MMLU-speech, while Mel-LLM drops from 58.68 to 41.30. We therefore view semantic anchoring and audio-to-reasoning alignment as the main bottleneck. Encoder-anchored systems can map audio into representations already aligned with the LLM's text reasoning space, while encoder-free systems must learn low-level acoustics and preserve reasoning behavior inside the same LLM. This improves access to paralinguistic and non-speech cues, but can weaken the path from speech perception to knowledge-intensive reasoning. Reducing this gap likely requires stronger native speech-text joint training, larger or better-preserved language backbones, and language-preservation objectives such as text replay or distillation during acoustic adaptation.

\subsection{Preliminary Speech Generation Results}

Table~\ref{tab:tts} reports zero-shot TTS results on LibriSpeech-PC test-clean. Direct continuous Mel generation with the LLM backbone is feasible only with Phi-4-MM initialization; random initialization converges but produces inaudible speech. However, the gap to the Phi-4-MM + latent diffusion baseline remains large. That baseline follows the VibeVoice-style next-token diffusion idea~\cite{peng2025vibevoice}, where the autoregressive model predicts latent representations and a diffusion decoder reconstructs speech. These results show that direct Mel generation is still exploratory: unlike speech understanding, high-fidelity generation must preserve linguistic content and acoustic naturalness while avoiding compounding frame-level errors.

\begin{table}[!t]
\centering
\caption{Zero-shot TTS results on LibriSpeech-PC test-clean~\cite{meister2023librispeechpc}. We follow the MELLE~\cite{meng2025melle} framework, autoregressively generating continuous mel frames via a VAE decoder. WER is measured by Whisper-large-v3~\cite{radford2023robust}; UTMOS~\cite{saeki2022utmos} measures perceptual quality. Results are preliminary.}
\vspace{-0.05cm}
\resizebox{\columnwidth}{!}{
\begin{tabular}{l|cc}
\toprule
System & WER $\downarrow$ & UTMOS $\uparrow$ \\
\midrule
Phi-4-MM + latent diffusion~\cite{peng2025vibevoice} & 4.2 & 3.41 \\
\midrule
Mel-LLM (Random init) & \multicolumn{2}{c}{converge but no audible output} \\
\midrule
Mel-LLM (Phi-4-MM, no norm) & 11.03 & 3.10 \\
Mel-LLM (Phi-4-MM, MVN) & 14.75 & 3.25 \\
\quad + dropout 0.1 & 85.51 & 1.38 \\
\quad + fix-KL (0-mean) & 12.65 & 3.22 \\
\quad + sigma-VAE (0-mean) & 18.07 & 3.29 \\
\bottomrule
\end{tabular}}
\label{tab:tts}
\vspace{-0.05cm}
\end{table}

\subsection{Scope and Practical Implications}

Encoder-free modeling does not remove the need for pretraining: Phi-4-MM LoRA initialization remains critical, and Mel-LLM retains lightweight downsampling and projection layers. Pretrained speech encoders remain effective and ubiquitous, and our results favor them for knowledge-intensive spoken QA, while latent-diffusion generation remains stronger for TTS. Mel-LLM should therefore not yet be viewed as a universal replacement for encoder-based Speech-LLMs. Its practical value lies in simplifying the audio stack, improving training efficiency, and preserving acoustic and paralinguistic information that a semantic encoder may compress. More broadly, our results identify conditions under which encoder removal is feasible and expose the resulting trade-off between acoustic access and semantic alignment.

\vspace{-0.05cm}
\section{Conclusion}
\label{sec:conclusion}
\vspace{-0.05cm}

We present Mel-LLM, demonstrating that LLM can directly learn to read Mel spectrogram without a dedicated speech encoder. On ASR, the encoder-free approach achieves competitive results with only limited performance gap compared to encoder-initialized models, particularly when sufficient training data is available. Phi-4-MM initialization proves critical for low-resource settings. Our ablation studies suggest that lower LLM layers are most relevant for implicit speech adaptation. Beyond ASR, Mel-LLM shows strong gains on paralinguistic and non-speech acoustic tasks such as emotion, gender, speaker, and music understanding, indicating that direct Mel access preserves information often compressed by semantic encoders. At the same time, weaker spoken-QA results reveal a key limitation: removing the dedicated encoder blocks also removes a semantic anchor, increasing the audio-to-reasoning bottleneck during acoustic adaptation. Future work should therefore focus on native speech-text joint training, text distillation to preserve LLM reasoning, auxiliary transcript or teacher-encoder supervision to anchor speech representations, and curriculum schedules that introduce acoustic modeling without overwhelming text-side performance. Our preliminary TTS experiments further suggest that direct continuous Mel generation with an LLM backbone remains substantially harder than speech understanding, motivating stronger latent generation heads for speech synthesis.

\section{Generative AI Use Disclosure}
We used AI tools for assitant writing, including grammar correction, spelling checks, formatting and drawing LaTeX tables. All technical content, experiments, and conclusions were produced and verified by the authors.

\end{document}